\documentclass[aps,prd,amsmath,twocolumn,amssymbaps,showpacs]{revtex4-1}
\usepackage{graphicx}  % needed for figures
\usepackage{dcolumn}   % needed for some tables
\usepackage{bm}        % for math
\usepackage{amssymb}   % for math
\usepackage{amsmath}   % for math
\usepackage{bbm}
\usepackage{draftcopy}
\usepackage{relsize} %  ! da scommentare alla fine
\usepackage{xfrac}    % ! da scommentare alla fine
\usepackage{slashed}
\usepackage{color}
\definecolor{dgreen}{cmyk}{1.,0.,1.,0.2}        % dark green
\definecolor{orange}{cmyk}{0.,0.353,1.,0.}    % orange

\usepackage[bookmarks]{hyperref}

\newcommand{\di}{{\rm d}}

\newcommand{\tr}{{\rm tr}}

\newcommand{\be}{\begin{equation}}
\newcommand{\ee}{\end{equation}}                                                                               
\newcommand{\bea}{\begin{eqnarray}}
\newcommand{\eea}{\end{eqnarray}}

\begin{document}
\title{Effects of imaginary and real rotations on QCD matters}
\author{Gaoqing Cao}
\affiliation{School of Physics and Astronomy, Sun Yat-sen University, Zhuhai 519088, China}
\date{\today}

\begin{abstract}
Inspired from perturbative calculations, this work introduces imaginary ($\Omega_{\rm I}$) and real ($\Omega$) rotation effects to the pure $SU(3)$ gauge potentials simply through variable transformations: The empirical Polyakov loop (PL) potentials can be rewritten as functions of the imaginary chemical potentials of gluons and ghosts $(q_{\rm ij})$, and the transformations are taken as $q_{\rm ij}\rightarrow q_{\rm ij}\pm\Omega_{\rm I}/T$ and  $q_{\rm ij}\rightarrow q_{\rm ij}\pm i\,\Omega/T$, respectively. For the PL potential of Fukushima $(V_1)$, a smaller imaginary rotation $\Omega_{\rm I}$ tends to suppress PL at all temperature and the deconfinement transition keeps of first order. However, for the PL potential of Munich group $(V_2)$, $\Omega_{\rm I}$ tends to enhance PL at low temperature $T$, consistent with lattice simulations; but suppress PL at high $T$, consistent with perturbative calculations. Moreover, the deconfinement alters from first order to crossover with increasing $\Omega_{\rm I}$ as is expected from lattice simulations. On the other hand, the real rotation $\Omega$ tends to enhance PL at relatively low $T$ for both potentials, and the (pseudo-)critical temperature decreases with $\Omega$ as expected. Therefore, we find that analytic continuation of the phase diagram from imaginary to real rotation is not necessarily valid in the non-perturbative region. Finally, we apply the more successful PL potential $V_2$ to the Polyakov--Nambu-Jona-Lasinio (PNJL) model and discover that $\Omega_{\rm I}$ tends to break chiral symmetry while $\Omega$ tends to restore it. Especially, the modified model is even able to qualitatively explain the lattice result that a larger $T$ would catalyze chiral symmetry breaking for a large $\Omega_{\rm I}$.
\end{abstract}

\maketitle
%%%%%%%%%%%%%%%%%%%%%%%%%%%%%%%%%%%%%%%%%%%%%%%%%%%%%%%%%%%%%%%%%%%%%%%%%%%%%%%%%%%%%%%%%%%%%%%%%%
\section{Introduction}
%%%%%%%%%%%%%%%%%%%%%%%%%%%%%%%%%%%%%%%%%%%%%%%%%%%%%%%%%%%%%%%%%%%%%%%%%%%%%%%%%%%%%%%%%%%%%%%%%%
Recently, rotational or vortical effects are of great interests to both theorists and experimentalists in high energy nuclear physics~\cite{Becattini:2021lfq}. Mainly, three aspects of rotation are extensively concerned in the literatures: the anomalous transport phenomena, the phase diagrams of QCD, and the polarizations of hyperons and $\phi$ mesons. The former two are more of theoretical interests and had been explored with a variety of methods or effective models, while the last closely connects theories to experiments and promotes mutual developments. The earliest study of vortex involved anomalous transport is chiral vortical effect~\cite{Son:2004tq,Metlitski:2005pr}, which was proposed right after the chiral magnetic effect~\cite{Kharzeev:2007jp,Fukushima:2008xe} inspired from the similar polarization effects of rotation and magnetic field. Later, more anomalous phenomena were discovered, such as chiral vortical separation effect~\cite{Landsteiner:2011iq}, magnetovorticity effect~\cite{Hattori:2016njk}, and chiral electric vortical effect~\cite{Cao:2021jjy,Yamamoto:2021gts}. In 2005, Liang and Wang proposed to search for the global polarization of hyperons in heavy ion collisions~\cite{Liang:2004ph} but no significant signal was detected in the high energy peripheral heavy ion collisions. The interests were renewed in 2016 when significant signals were found at smaller values of collision energy~\cite{STAR:2017ckg}, and the local polarizations were explored for the first time in both the longitudinal and transversal directions~\cite{Becattini:2017gcx,Xia:2018tes,Niida:2018hfw,Becattini:2019ntv,Xia:2019fjf,Florkowski:2019voj,Cao:2021nhi,Becattini:2021iol,Fu:2021pok}.

If one translates the polarization signals into the angular velocities of rotation, the magnitude was evaluated to be as large as $\Omega=9\times 10^{21} s^{-1}$~\cite{STAR:2017ckg}, thus the quark-gluon plasma was regarded as the most vortical fluid in the universe. In natural units, the rotation velocity is $\Omega=6~{\rm MeV}$, and the effective chemical potential $l\,\Omega$ is comparable to the QCD scale $\Lambda_{\rm QCD}\sim200~{\rm MeV}$ for the angular momentum $l\geq 30$. Since then, the effects of rotation on QCD phases were extensively studied including the traditional topics such as chiral symmetry and confinement~\cite{Jiang:2016wvv,Chen:2015hfc,Ebihara:2016fwa}, and the possibilities of color superconductivity~\cite{Jiang:2016wvv}, pion superfluidity~\cite{Liu:2017spl,Cao:2019ctl,Chen:2019tcp} and rho meson superconductivity~\cite{Zhang:2018ome,Cao:2020pmm}. The first studies of such rotational effects were carried out in the chiral effective NJL model where quarks are the elementary degrees of freedom~\cite{Jiang:2016wvv,Chen:2015hfc,Ebihara:2016fwa}. It was found that rotation tends to suppress pairing states with zero angular momentum, such as chiral condensates and $2{\rm SC}$ diquark condensates~\cite{Jiang:2016wvv}. Later, the Klein-Gordon theory predicted that pion superfluidity could appear in a QCD system with parallel magnetic field and rotation~\cite{Liu:2017spl}, which was then confirmed by the studies in the NJL model~\cite{Cao:2019ctl,Chen:2019tcp}. Furthermore, such a circumstance was carefully checked in advance and rho meson superconductivity was found to be more favored for a larger magnetic field~\cite{Cao:2020pmm}. 

These years, several lattice QCD (LQCD) simulations were carried out to understand the features of confinement and chiral symmetry in the presence of an imaginary rotation $\Omega_{\rm I}$ as there is no sign problem~\cite{Braguta:2021jgn,Chernodub:2022veq,Yang:2023vsw}. There seem contradictions among the results: For homogeneous phases, $\Omega_{\rm I}$ would break confinement and restore chiral symmetry, hence the pseudocritical temperature $T_{\rm c}$ decreases with $\Omega_{\rm I}$ for a finite system~\cite{Braguta:2021jgn,Yang:2023vsw}. For inhomogeneous phases, $\Omega_{\rm I}$ would suppress deconfinement and $T_{\rm c}$ increases with $\Omega_{\rm I}$~\cite{Chernodub:2022veq}. In the case of a real rotation, there is sign problem in principle as the effective chemical potentials $l\,\Omega$ would render the action complex valued~\cite{Chen:2015hfc}. So, one usually supposes the analytic continuation to be valid for finite rotations and obtains the phase diagram of $\Omega$ from that of  $\Omega_{\rm I}$. Then, if one translates the homogeneous results, the conclusion would be that $T_{\rm c}$ increases with $\Omega$~\cite{Braguta:2021jgn,Yang:2023vsw}, opposite to the findings in the NJL model~\cite{Jiang:2016wvv,Chen:2015hfc,Ebihara:2016fwa}. To understand that, perturbative calculations and effective models had been applied to such systems but most of the results turned out to be "unsuccessful"~\cite{Chen:2022smf,Chen:2020ath,Chen:2022mhf,Chen:2023cjt} except few studies~\cite{Jiang:2021izj,Mameda:2023sst}. Especially, the perturbative calculations started with the completely deconfined phase at high temperature, where the PL is $1$ and only gluons are the relevant degrees of freedom, and found that $\Omega_{\rm I}$ would reinforce confinement~\cite{Chen:2022smf}. However, ghosts are the reasons of confinement at low temperature in the language of LQCD~\cite{Fukushima:2017csk} and would also be directly affected by rotations~\cite{Chen:2022smf}. In that sense, the perturbative results do not necessarily contradict with the LQCD simulations, where the PL was found to be much less than $1$ around $T_{\rm c}$~\cite{Fukushima:2017csk}.

In this work,  we want to find a way to reasonably introduce rotation effects into the gluon sector of QCD and then check whether it is reliable or not to obtain real rotation effect simply through analytic continuation of the LQCD data. The paper is organized as follows. Within Sec.\ref{PSGT}, we try to introduce the effects of imaginary and real rotations into the pure $SU(3)$ gauge theory in Sec.\ref{RE} and compare the results of two empirical Polyakov loop potentials in Secs.\ref{V1} and \ref{V2}, respectively. Then, we extend the work to the three-flavor PNJL model to study the properties of QCD system more realistically in Sec.\ref{3PNJL}. Finally, a summary will be given in Sec.\ref{sum}.

%%%%%%%%%%%%%%%%%%%%%%%%%%%%%%%%%%%%%%%%%%%%%%%%%%%%%%%%%%%%%%%%%%%%%%%%%%%%%%%%%%%%%%%%%%%%%%%%%%
\section{The pure $SU(3)$ gauge theory}\label{PSGT}
%%%%%%%%%%%%%%%%%%%%%%%%%%%%%%%%%%%%%%%%%%%%%%%%%%%%%%%%%%%%%%%%%%%%%%%%%%%%%%%%%%%%%%%%%%%%%%%%%%
%%%%%%%%%%%%%%%%%%%%%%%%%%%%%%%%%%%%%%%%%%%%%%%%%%%%%%%%%%%%%%%%%%%%%%%%%%%%%%%%%%%%%%%%%%%%%%%%%%
\subsection{Introduction of rotation effects}\label{RE}
%%%%%%%%%%%%%%%%%%%%%%%%%%%%%%%%%%%%%%%%%%%%%%%%%%%%%%%%%%%%%%%%%%%%%%%%%%%%%%%%%%%%%%%%%%%%%%%%%%
In the pure $SU(3)$ gauge theory, Polyakov loop serves as a true order parameter for the $Z_3$ center symmetry of the gauge group, which is directly related to confinement~\cite{Fukushima:2017csk}.  It is defined as $L\equiv{1\over N_{\rm c}}\tr_{\rm c}\ e^{i\,g\int_0^\beta{\cal A}_4\di\tau}$ where $\beta=1/T$ is inverse temperature and ${\cal A}_4\equiv A_4^{a}{\lambda^{a}_{\rm c}\over2}$ is the temporal component of the non-Abelian gauge field with $\lambda^a_{\rm c}\ (a=1-8)$ the Gell-Mann matrices in color space. Usually, the expectation value of ${\cal A}_4$ is taken as a diagonal and traceless constant matrix, that is, $ g{\cal A}_4=T\,{\rm diag}(q_1,q_2,q_3)$ with $q_1+q_2+q_3=0$~\cite{Fukushima:2017csk}; then it follows that 
\bea
L(q_1,q_2,q_3)={1\over N_{\rm c}}(e^{i\, q_1}+e^{i\,q_2}+e^{i\, q_3}).\label{PL}
\eea
With respect to that, imaginary color chemical potentials are introduced to both gluons and ghosts, that is,
 \bea
\pm q_{31}, \pm q_{21},\pm q_{32}
\eea
with $q_{\rm ij}=q_i-q_j$~\cite{Fukushima:2017csk}. Since $\pm q_{\rm ij}$ always show up together, we only take $q_{\rm ij}\ (i>j)$ as the independent variables in the following. 

\begin{figure}[!htb]
	\begin{center}
	\includegraphics[width=8cm]{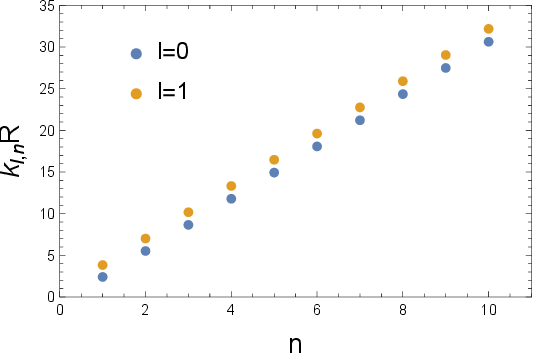}
		\caption{The $n$-th eigenenergies $k_{l,n}$ that satisfy $J_l(k_{l,n} R)=0$ for $l=0$ (blue) and $l=1$ (yellow).}\label{Jl}
	\end{center}
\end{figure}

According to the perturbative study in Ref.~\cite{Chen:2022smf}, the imaginary rotation ${\Omega}_{\rm I}$ introduces extra imaginary chemical potentials to both gluons and ghosts. And in the deconfined phase, the effect can be equivalently accounted for by simply taking two branches of variable transformations to $q_{\rm ij}$ in the thermodynamic potential of physical gluons, that is,
\bea
q_{\rm ij}&\rightarrow&q_{\rm ij}+s\,\tilde{\Omega}_{\rm I}\label{trans}
\eea
with $\tilde{\Omega}_{\rm I}\equiv{\Omega_{\rm I}\over T}$, and averaging over $s=\pm1$. On the other hand, the contributions of ghosts were found to be indispensable in order to explain confinement at low temperature~\cite{Fukushima:2017csk}. How should the imaginary rotation effect be introduced to the ghosts? In the deconfined phase, ${\Omega}_{\rm I}$ would affect ghosts in a similar way as physical gluons; but to introduce an effective chemical potential, $i\, l\, \Omega_{\rm l}$, states with orbital angular momentum (OAM) $l$ are required for the scalar ghosts while the ones with OAM $l\pm1$ are required for the transversal gluons~\cite{Chen:2022smf}. For the origin $r=0$~\cite{Chen:2022smf}, it is true that ${\Omega}_{\rm I}$ does not affect the ghosts as only the states with zero $l$ contribute. However, lattice QCD usually explore the spatial average of PL~\cite{Braguta:2021jgn,Yang:2023vsw}, in which case the contributions of the states with $l\neq 0$ could also be important. Take a cylindrical system with radius $R$ for example,  we compare the $n$-th eigenenergies $k_{0,n}$ and $k_{1,n}$ that satisfy the boundary condition $J_l(k_{l,n} R)=0$ in Fig.~\ref{Jl}.
As can be seen, $k_{1,n}$ is slightly larger than $k_{0,n}$ for a given $n$. Actually, there are more restrict unequal relations among the eigenenergies, that is,~\cite{Abramowitz1964}
\bea
k_{0,n}<k_{1,n}<k_{0,n+1}<k_{1,n+1};
\eea
so $\Omega_{\rm l}$ could induce remarkable effect to ghosts on average. In light of that, we will intuitively introduce the rotational effect to gluons and ghosts by considering the leading-order nontrivial contributions, that is, the transformations in Eq.\eqref{trans} would be extended to the thermodynamic potential of confined phase.

For the pure $SU(3)$ gauge theory, the thermodynamic potential is usually presented as a function of PL with the expressions obtained by fitting to lattice simulations~\cite{Fukushima:2017csk}. The advantage of such a thermodynamic potential is that the PL can be self-consistently calculated from the gap equation thus the confinement-deconfinement transition can be well identified from the feature of the order parameter. In the market, there are two most popular parameterizations of the PL potential $V$, that is,
\bea
{V_1(L,L^*)\over T^4}&=&-{a(\tilde{T})\over2}|L|^2-{1.75\over\tilde{T}^{3}}\ln H(L,L^*),\\
&&\!\!\!\!\!\!\!\!\!\!\!\!\!\!\!\!\!\!\!\!\!\!\!\!\!\!\!a(\tilde{T})=3.51-{2.47\over\tilde{T}}+{15.2\over\tilde{T}^{2}},\nonumber
\eea
given by K. Fukushima~\cite{Fukushima:2003fw} and
\bea
\!\!\!\!\!\!\!\!{V_2(L,L^*)\over T^4}&=&-{a(\tilde{T})\over2}|L|^2-{0.75\over6}\left(L^3+{L^*}^3\right)+{7.5\over4}|L|^4,\\
&&a(\tilde{T})=6.75-{1.95\over\tilde{T}}+{2.625\over\tilde{T}^{2}}-{7.44\over\tilde{T}^{3}}\nonumber
\eea
given by Munich group~\cite{Ratti:2005jh}. Here, $\tilde{T}\equiv T/T_0$ is the reduced temperature with $T_0=0.27\,{\rm GeV}$, and the Haar measure is defined as~\cite{Fukushima:2017csk}
\bea
H(L,L^*)&=&27\Big[1\!-\!6|L|^2\!+\!4(L^3\!+\!{L^*}^3)-3|L|^4\Big]\nonumber\\
&=&\prod_{\rm i,j=1,2,3; i>j}\left|e^{i\, q_{\rm i}}-e^{i\, q_{\rm j}}\right|^2.
\eea

In order to effectively introduce rotational effect into the PL potentials, all the terms involved in $V_1$ and $V_{2}$ must be re-expressed as functions of $q_{\rm ij}$. The relevant terms can be rewritten as
 \bea
|L|^2(q_{\rm ij})&=&{1\over N_{\rm c}}+{1\over N_{\rm c}^2}\sum_{t=\pm}(e^{t\,i\, q_{31}}+e^{t\,i\,q_{21}}+e^{t\,i\, q_{32}}),\\
H(q_{\rm ij})&=&\prod_{ i,j=1,2,3; i>j}(2\!-\!e^{i\, q_{\rm ij}}\!-\!e^{-i\, q_{\rm ij}}),\\
L^3(q_{\rm ij})&=&{1\over N_{\rm c}^3}e^{3i\, q_1}\left(1+e^{i\,q_{21}}+e^{i\, q_{31}}\right)^3\nonumber\\
&=&{1\over N_{\rm c}^3}e^{-i\, (q_{21}+q_{31})}\left(1+e^{i\,q_{21}}+e^{i\, q_{31}}\right)^3,\\
{L^*}^3(q_{\rm ij})&=&{1\over N_{\rm c}^3}e^{i\, (q_{21}+q_{31})}\left(1+e^{-i\,q_{21}}+e^{-i\, q_{31}}\right)^3,
\eea
then the PL potentials become
 \bea
{V_1(q_{\rm ij})\over T^4}&=&-{a(\tilde{T})\over2}|L|^2(q_{\rm ij})-{1.75\over\tilde{T}^{3}}\ln H(q_{\rm ij}),\\
{V_2(q_{\rm ij})\over T^4}&=&-{a(\tilde{T})\over2}|L|^2(q_{\rm ij})-{0.75\over6}\left[L^3(q_{\rm ij})+{L^*}^3(q_{\rm ij})\right]\nonumber\\
&&+{7.5\over4}\left[|L|^2(q_{\rm ij})\right]^2.
\eea
So, by applying the variable transformations in Eq.\eqref{trans}, the PL potentials with imaginary rotation effect are, respectively,
 \bea
{\bar{V}_{1}(q_{\rm ij},\Omega_{\rm I})}&=&{1\over2}\sum_{s=\pm}V_{1}(q_{\rm ij}+s\,\Omega_{\rm I}),\\
{\bar{V}_{2}(q_{\rm ij},\Omega_{\rm I})}&=&{1\over2}\sum_{s=\pm}V_{2}(q_{\rm ij}+s\,\Omega_{\rm I}).
\eea
Eventually, recalling that in Ref.~\cite{Chen:2022smf} the color chemical potentials were alternatively expressed as
\bea
q_{31}=\phi_1,\
q_{21}={\phi_1\over2}+{\sqrt{3}\over2}\phi_2,\
q_{32}={\phi_1\over2}-{\sqrt{3}\over2}\phi_2\label{qij}
\eea
with $\phi_1\in[0,2\pi]$ and $\phi_2\in[-{\phi_1\over\sqrt{3}},{\phi_1\over\sqrt{3}}]$, we can further rewrite the PL potentials as functions of $\phi_1$ and $\phi_2$, that is,
\bea
\!\!\!\!\!\!\!\!\!\!\!V_{1}(\phi_1,\phi_2,\Omega_{\rm I})&=&\bar{V}_{1}(\phi_1,{\phi_1\over2}\!+\!{\sqrt{3}\over2}\phi_2,{\phi_1\over2}\!-\!{\sqrt{3}\over2}\phi_2,\Omega_{\rm I}),\label{V11}\\
\!\!\!\!\!\!\!\!\!\!\!V_{2}(\phi_1,\phi_2,\Omega_{\rm I})&=&\bar{V}_{2}(\phi_1,{\phi_1\over2}\!+\!{\sqrt{3}\over2}\phi_2,{\phi_1\over2}\!-\!{\sqrt{3}\over2}\phi_2,\Omega_{\rm I}).\label{V21}
\eea

One can check that $V(\phi_1,\phi_2,\Omega_{\rm I})$ is real valued as should be, since there is no sign problem for the QCD action with finite $\Omega_{\rm I}$. Thus, we can search for the global minimum of the thermodynamic potentials with respect to $\phi_1$ and $\phi_2$ in the constrained region, and then the PL $L$ can be evaluated according to Eq.\eqref{PL}. For the latter purpose, we work out $q_{\rm i}\ (i=1,2,3)$ first as functions of $\phi_1$ and $\phi_2$ according to Eq.\eqref{qij}, and we have
\bea
q_1=-{\phi_1\over2}-{\phi_2\over2\sqrt{3}},\ q_2={\phi_2\over\sqrt{3}},\ q_3={\phi_1\over2}-{\phi_2\over2\sqrt{3}}.\label{qi}
\eea
Note that $L$ is not necessarily real-valued at finite $\Omega_{\rm I}$, so we will eventually present the absolute value of $L$ instead.
To determine $\phi_1$ and $\phi_2$, gap equations can be derived by following $\partial V/\partial \phi_1=\partial V/\partial \phi_2=0$, that is,
\begin{widetext}
\bea
\!\!\!\!0&=&\!-{i\,a(\tilde{T})\over8N_{\rm c}^2}\!\!\sum_{t,s=\pm}\!t\left[2e^{t\,i\, q_{31}^s}\!+\!e^{t\,i\,q_{21}^s}\!+\!e^{t\,i\, q_{32}^s}\right]\!+\!{1.75i\over4\tilde{T}^{3}}\!\sum_{s=\pm}\!\!\left[{2(e^{i\, q_{31}^s}\!-\!e^{-i\, q_{31}^s})\over 2\!-\!e^{i\, q_{31}^s}\!-\!e^{-i\, q_{31}^s}}\!+\!{e^{i\, q_{21}^s}\!-\!e^{-i\, q_{21}^s}\over 2\!-\!e^{i\, q_{21}^s}\!-\!e^{-i\, q_{21}^s}}\!+\!{e^{i\, q_{32}^s}\!-\!e^{-i\, q_{32}^s}\over 2\!-\!e^{i\, q_{32}^s}\!-\!e^{-i\, q_{32}^s}}\right],\\
\!\!\!\!0&=&\!-{\sqrt{3}i\,a(\tilde{T})\over8N_{\rm c}^2}\sum_{t,s=\pm}t\,\left(e^{t\,i\,q_{21}^s}-e^{t\,i\, q_{32}^s}\right)+{1.75\sqrt{3}i\over4\tilde{T}^{3}}\sum_{s=\pm}\left({e^{i\, q_{21}^s}\!-\!e^{-i\, q_{21}^s}\over 2\!-\!e^{i\, q_{21}^s}\!-\!e^{-i\, q_{21}^s}}-{e^{i\, q_{32}^s}\!-\!e^{-i\, q_{32}^s}\over 2\!-\!e^{i\, q_{32}^s}\!-\!e^{-i\, q_{32}^s}}\right)
\eea
for $V_1(\phi_1,\phi_2,\Omega_{\rm I})$ with $q_{\rm ij}^s=q_{\rm ij}+s\,\Omega_{\rm I}$ and 
\bea
0&=&\sum_{t,s=\pm}\Bigg\{-{i\,a(\tilde{T})\over8N_{\rm c}^2}t\,\left[2e^{t\,i\, q_{31}^s}+e^{t\,i\,q_{21}^s}+e^{t\,i\, q_{32}^s}\right]+{0.75i\over8N_{\rm c}^3}
t\,\left[e^{-t\,i\, (q_{21}^s+q_{31}^s)}\left(1+e^{t\,i\,q_{21}^s}+e^{t\,i\, q_{31}^s}\right)^2(1-e^{t\,i\, q_{31}^s})\right]\nonumber\\
&&+{7.5i\over4N_{\rm c}^2}t\,\left(e^{t\,i\, q_{31}^s}+{1\over 2}e^{t\,i\,q_{21}^s}+{1\over 2}e^{t\,i\, q_{32}^s}\right)|L|^2(q_{\rm ij}^{s})\Bigg\},\label{dV11}\\
0&=&\sum_{t,s=\pm}\left\{-{\sqrt{3}i\,a(\tilde{T})\over8N_{\rm c}^2}t\,\left(e^{t\,i\,q_{21}^s}-e^{t\,i\, q_{32}^s}\right)+{0.75\sqrt{3}i\over24N_{\rm c}^3}t\,
\left[e^{-t\,i\, (q_{21}^s+q_{31}^s)}\left(1+e^{t\,i\,q_{21}^s}+e^{t\,i\, q_{31}^s}\right)^2(1+e^{t\,i\, q_{31}^s}-2e^{t\,i\,q_{21}^s})\right]\right.\nonumber\\
&&\left.+{7.5\sqrt{3}i\over8N_{\rm c}^2}t\,\left(e^{t\,i\,q_{21}^s}-e^{t\,i\, q_{32}^s}\right)|L|^2(q_{\rm ij}^{s})\right\}\label{dV12}
\eea
for $V_2(\phi_1,\phi_2,\Omega_{\rm I})$.
\end{widetext}

Finally, the formalism with real rotation $\Omega$ can be obtained from the one with imaginary rotation $\Omega_{\rm I}$ by taking the analytic continuation: $\Omega_{\rm I}\rightarrow-i\,\Omega$. One can easily check that the thermodynamic potential $V(\phi_1,\phi_2,-i\Omega)$ is still real valued, so we can pin down the ground state by minimizing $V(\phi_1,\phi_2,-i\Omega)$ over $\phi_1$ and $\phi_2$. It is easy to see from the exponentials 
\bea
e^{t\,i\, q_{\rm ij}^s}\rightarrow e^{t\,(i\, q_{\rm ij}+s\,\Omega)}
\eea
 that real rotation $\Omega$ functions as a real chemical potential to gluons and ghosts. Usually, gluons tends to break confinement while ghosts tends to reinforce it in the language of lattice QCD. So recalling that the contributions of ghosts are larger than those of gluons at low temperature~\cite{Fukushima:2017csk}, one might expect that $\Omega$ favors confinement when both numbers of gluons and ghosts increase with $\Omega$. As mentioned in Ref.~\cite{Ebihara:2016fwa}, boundary effect has to be taken into account self-consistently for real rotation in order to satisfy the causality. Here, since there is no summation over transversal eigenenergy in the empirical PL potentials, we simply neglect that.

%%%%%%%%%%%%%%%%%%%%%%%%%%%%%%%%%%%%%%%%%%%%%%%%%%%%%%%%%%%%%%%%%%%%%%%%%%%%%%%%%%%%%%%%%%%%%%%%%%
\subsection{Numerical results}
%%%%%%%%%%%%%%%%%%%%%%%%%%%%%%%%%%%%%%%%%%%%%%%%%%%%%%%%%%%%%%%%%%%%%%%%%%%%%%%%%%%%%%%%%%%%%%%%%%
For the Polyakov loop potentials $V_1(\phi_1,\phi_2,\Omega_{\rm I})$ and $V_2(\phi_1,\phi_2,\Omega_{\rm I})$ just developed, the corresponding numerical results are presented in Sec.\ref{V1} and Sec.\ref{V2}, respectively.

%%%%%%%%%%%%%%%%%%%%%%%%%%%%%%%%%%%%%%%%%%%%%%%%%%%%%%%%%%%%%%%%%%%%%%%%%%%%%%%%%%%%%%%%%%%%%%%%%%
\subsubsection{Polyakov loop potential $V_1$}\label{V1}
%%%%%%%%%%%%%%%%%%%%%%%%%%%%%%%%%%%%%%%%%%%%%%%%%%%%%%%%%%%%%%%%%%%%%%%%%%%%%%%%%%%%%%%%%%%%%%%%%%
\begin{figure}[!htb]
	\begin{center}
	\includegraphics[width=8cm]{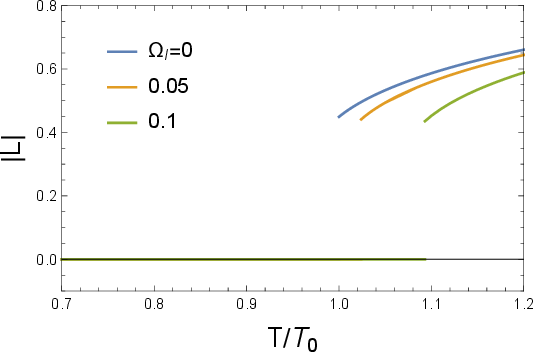}
		\caption{The absolute value of Polyakov loop, $|L|$, as a function of temperature $T$ for imaginary rotations $\Omega_{\rm I}=0, 0.05$ and $0.1~{\rm GeV}$.}\label{PL2I}
	\end{center}
\end{figure}
\begin{figure}[!htb]
	\begin{center}
	\includegraphics[width=8cm]{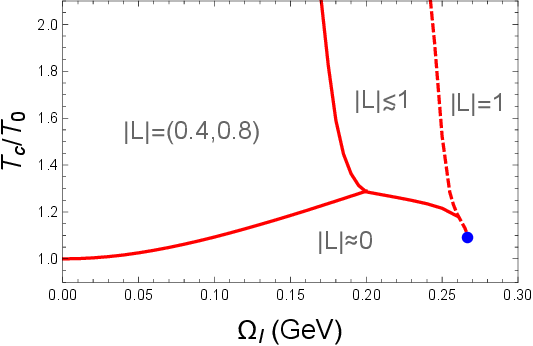}
		\caption{The temperature-imaginary rotations ($T-\Omega_{\rm I}$) phase diagram. The solid and dashed lines correspond to first- and second-order transitions, respectively, and the blue bullet is a critical end point.}\label{CT2I}
	\end{center}
\end{figure}

In Fig.~\ref{PL2I}, we demonstrate the PL $|L|$ as a function of temperature $T$ for several imaginary rotations. As can be seen, $\Omega_{\rm I}$ tends to suppress $|L|$ for all $T$ and the deconfinement transition is of first order -- both features are consistent with the predictions of perturbative calculations~\cite{Fukushima:2017csk}. Remember that the transition is of strong first order at $\Omega_{\rm I}=0$ for $V_1$~\cite{Fukushima:2003fw}, so it is not surprising that it remains of first order even when $\Omega_{\rm I}$ is large. In fact, the imaginary rotation effect is quite nontrivial in the Haar measure term and lots of local minima could be developed with respect to $\phi_1$ and $\phi_ 2$ at mediate $T$, hence several branches of transitions can be involved. For a wider range of $\Omega_{\rm I}$, the relevant transition temperature is illustrated in Fig.~\ref{CT2I}, where three branches of transitions can be identified: $|L|\approx0\rightarrow |L|\approx0.4$, $|L|\approx0.8\rightarrow |L|\lesssim 1$, and $|L|\lesssim 1\rightarrow |L|=1$. Though the critical temperature increases with $\Omega_{\rm I}$ for the deconfinement transition $|L|\approx0\rightarrow |L|\approx0.4$, as shown in Fig.~\ref{PL2I}, it decreases with $\Omega_{\rm I}$ for the the other transitions and there is even a critical end point.

\begin{figure}[!htb]
	\begin{center}
	\includegraphics[width=8cm]{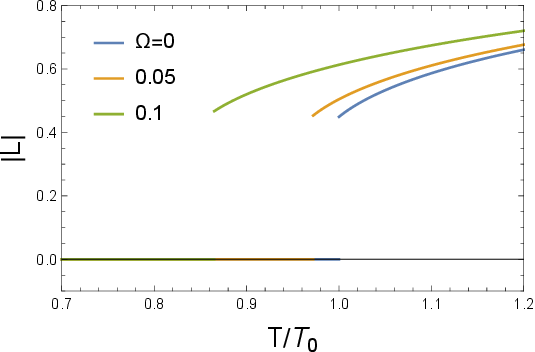}
		\caption{The absolute value of Polyakov loop, $|L|$, as a function of temperature $T$ for real rotations $\Omega=0, 0.05$ and $0.1~{\rm GeV}$.}\label{PL2R}
	\end{center}
\end{figure}

\begin{figure}[!htb]
	\begin{center}
	\includegraphics[width=8cm]{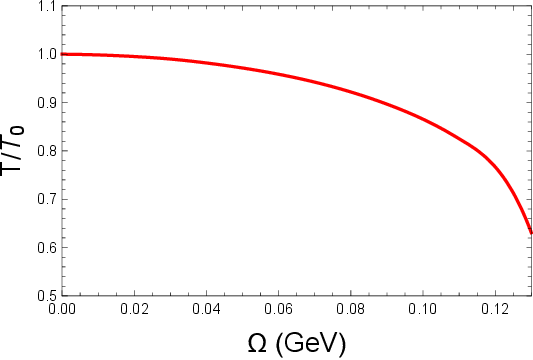}
		\caption{The temperature-real rotations ($T-\Omega$) phase diagram with the transition of first order.}\label{CT2R}
	\end{center}
\end{figure}

For the case with real rotation $\Omega$, we demonstrate the PL $|L|$ as a function of temperature $T$ in Fig.~\ref{PL2R} and the critical temperature as a function of $\Omega$ in Fig.~\ref{CT2R}, respectively. Comparing Fig.~\ref{PL2R} with Fig.~\ref{PL2I}, we find that real rotation tends to break confinement while imaginary rotation tends to reinforce it, as was shown in the perturbative study~\cite{Chen:2022smf}. Such an effect of real rotation is opposite to the naive analysis in the end of the previous section and implies that the centrifugal splitting effect overcomes the particle number enhancing effect. Moreover, the mathematical structure of $V_{1}(\phi_1,\phi_2,-i\,\Omega)$ as a function of $\phi_1$ and $\phi_ 2$ is much simpler than that of $V_{1}(\phi_1,\phi_2,\Omega_{\rm I})$ for given rotations, so only one branch of first-order phase transition shows up in Fig.~\ref{CT2R}. Since this branch corresponds to the deconfinement transition $|L|\approx0\rightarrow |L|\approx0.4$ in Fig.~\ref{PL2I}, analytic continuation of the phase digram from imaginary to real rotations seems to be valid here. 

%%%%%%%%%%%%%%%%%%%%%%%%%%%%%%%%%%%%%%%%%%%%%%%%%%%%%%%%%%%%%%%%%%%%%%%%%%%%%%%%%%%%%%%%%%%%%%%%%%
\subsubsection{Polyakov loop potential $V_2$}\label{V2}
%%%%%%%%%%%%%%%%%%%%%%%%%%%%%%%%%%%%%%%%%%%%%%%%%%%%%%%%%%%%%%%%%%%%%%%%%%%%%%%%%%%%%%%%%%%%%%%%%%
\begin{figure}[!htb]
	\begin{center}
	\includegraphics[width=8cm]{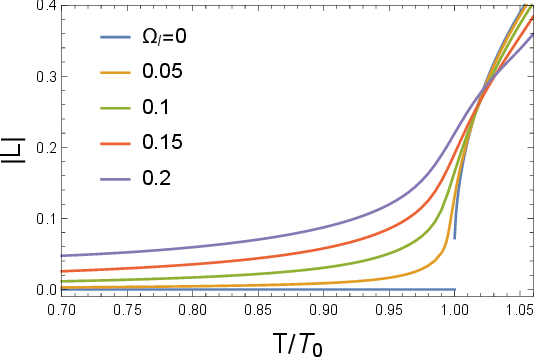}
		\caption{The absolute value of Polyakov loop, $|L|$, as a function of temperature $T$ for imaginary rotations $\Omega_{\rm I}=0, 0.05,0.1,0.15$ and $0.2~{\rm GeV}$.}\label{PL1I}
	\end{center}
\end{figure}
\begin{figure}[!htb]
	\begin{center}
	\includegraphics[width=8cm]{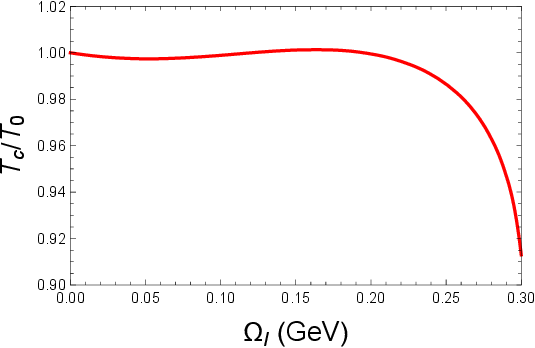}
		\caption{The temperature-imaginary rotations ($T-\Omega_{\rm I}$) phase diagram with the transition of crossover except for the region with $\Omega_{\rm I}\sim0$.}\label{CT1I}
	\end{center}
\end{figure}

We demonstrate the PL $|L|$ as a function of temperature $T$ for different imaginary rotations in Fig.~\ref{PL1I}. As can be seen, $\Omega_{\rm I}$ tends to enhance $|L|$ at a relatively small $T$, consistent with lattice simulations~\cite{Braguta:2021jgn,Yang:2023vsw}; but suppress it at a relatively large $T$, consistent with perturbative calculations~\cite{Fukushima:2017csk}. In fact, the sign change of $a(\tilde{T})$ around $T=T_0$ is responsible for such anomalous features: For $T$ a bit larger than $T_0$, $a(\tilde{T})>0$ and $\Omega_{\rm I}$ reinforces confinement, as had been shown in Ref.~\cite{Fukushima:2017csk}; for $T\lessapprox T_0$, $a(\tilde{T})<0$ and $\Omega_{\rm I}$ breaks confinement. The consistent reverse of imaginary rotation effect around $T\sim T_0$ is a strong support of the explanation. Since $a(\tilde{T})>0$ for all $T$ in $V_1$, such a reverse does not show up at all in Fig.~\ref{PL2I}. Another important observation from Fig.~\ref{PL1I} is that the transition is of weak first-order at $\Omega_{\rm I}=0$ but becomes crossover for a relatively large $\Omega_{\rm I}$. We can define the pseudocritical temperature $T_{\rm c}$ by the peak of $\partial_T |L|$, and the $T-\Omega_{\rm I}$ phase diagram is illustrated in Fig.~\ref{CT1I}. Though there is a small oscillation in the phase diagram, the main trend is that $T_{\rm c}$ decreases with $\Omega_{\rm I}$ which is consistent with lattice simulations~\cite{Braguta:2021jgn,Yang:2023vsw}. If the finite boundary effect is self-consistently taken into account, we believe the oscillation would eventually disappear as the transition is smoothed out around $T=T_0$~\cite{Braguta:2021jgn}. In that case, the reverse may also disappear and the results become more consistent with lattice simulations.
 
\begin{figure}[!htb]
	\begin{center}
	\includegraphics[width=8cm]{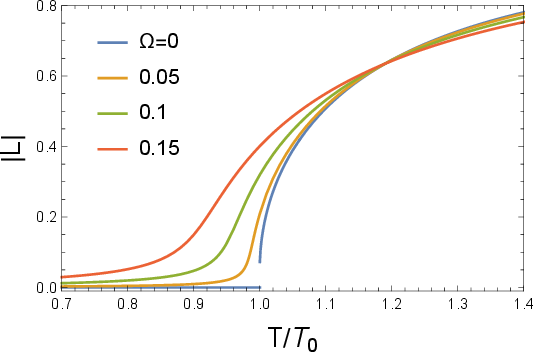}
		\caption{The absolute value of Polyakov loop, $|L|$, as a function of temperature $T$ for real rotations $\Omega=0, 0.05,0.1$ and $0.15~{\rm GeV}$.}\label{PL1R}
	\end{center}
\end{figure}

\begin{figure}[!htb]
	\begin{center}
	\includegraphics[width=8cm]{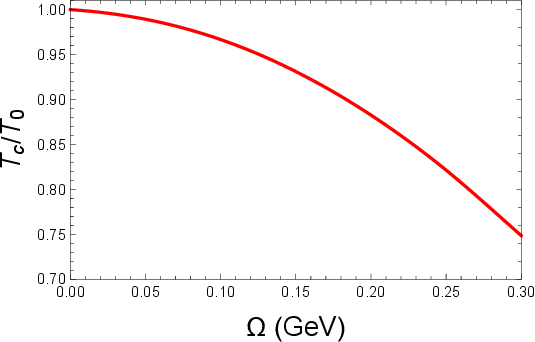}
		\caption{The temperature-real rotations ($T-\Omega$) phase diagram with the transition of crossover except for the region with $\Omega\sim0$.}\label{CT1R}
	\end{center}
\end{figure}

For the case with real rotation $\Omega$, we demonstrate the PL $|L|$ as a function of temperature $T$ in Fig.~\ref{PL1R} and the (pseudo-)critical temperature as a function of $\Omega$ in Figs.~\ref{CT1R}, respectively. Contrary to the findings in Sec.\ref{V1}, here the effect of real rotation is the same as that of imaginary rotation on confinement, compare Fig.~\ref{PL1R} with Fig.~\ref{PL1I} and Figs.~\ref{CT1R} with Figs.~\ref{CT1I}. Still, the opposite features of $|L|$ for $T<1.2\,T_0$ and $T>1.2\,T_0$ in Fig.~\ref{PL1R} are reflections of sign change of $a(\tilde{T})$ around $T_0$. And the study in this section provides an example that the analytic continuation of the phase diagram from imaginary to real rotation breaks down. On the other hand, it is interesting that even though the effect of imaginary rotation diverges from one PL potential to another one, the effect of real rotation are qualitatively the same. The consistency of real rotation effect is meaningful since that is the physical situation we are really interested in.

\begin{figure}[!htb]
	\begin{center}
	\includegraphics[width=8cm]{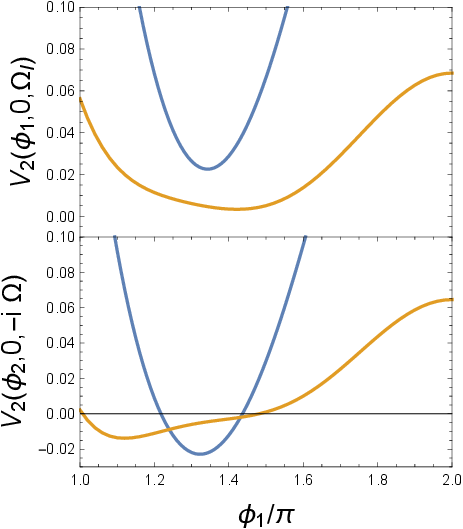}
		\caption{Assume $\phi_2=0$, the potential $V_2$ is demonstrated as a function of $\phi_1$ for $\Omega_{\rm I}=0.05~{\rm GeV}$ (upper panel) and $\Omega=0.05~{\rm GeV}$ (lower panel) with $T/T_0=0.9$ (blue) and $1$ (yellow).}\label{V1IR}
	\end{center}
\end{figure}

Why should the effects of imaginary and real rotations be the same whence the second order contributions of $\Omega_{\rm I}$ and  $\Omega$ are nonzero for $V_2$? The point is that the PL potential is not a trivial function of $L$ and $L^*$ anymore when $\Omega_{\rm I}$ or $\Omega$ is introduced; rather, it depends on them through the angles $\phi_1$ and $\phi_2$. Take $\phi_2=0$, which is true for the no rotation case, for example, the potential $V_2$ is demonstrated as a function of $\phi_1$ for $\Omega_{\rm I}=\Omega=0.05~{\rm GeV}$ and $T/T_0=0.9, 1$ in Fig.~\ref{V1IR}. Though the minima are both closely located around $\phi_1={4\pi\over3}$ for $T/T_0=0.9$, the temperature $T=T_0$ tends to shift the minimum to larger $\phi_1$ for the imaginary rotation while to smaller $\phi_1$ for the real rotation. In this sense, the effects of $\Omega_{\rm I}$ and $\Omega$ are indeed opposite to each other. However, recalling that $L=0$ for $\phi_1={4\pi\over3}$, both shifts tend to enhance the absolute value of PL, $|L|=|{1\over3}+{2\over3} \cos {\phi_1\over2}|$ -- that is exactly what we found in Figs.~\ref{PL1I} and \ref{PL1R}. Nevertheless, we will see in Sec.\ref{3PNJL} that the two trends would have different consequences on chiral symmetry breaking and restoration when quarks are taken into account.

We summarize the main results about the effects of imaginary ($\Omega_I$) and real ($\Omega$) rotations found from the Polyakov loop potentials $V_1$ and $V_2$ in Table.~\ref{V12} together with LQCD results.
\begin{table}[!h]
\centering
\begin{tabular}{|m{1cm}<{\centering}|m{2cm}<{\centering}|m{2cm}<{\centering}|m{2cm}<{\centering}|}
\hline
 & $V_1$ & $V_2$ & LQCD \\
 \hline\hline
$\Omega_I$ & $T_c\uparrow$ \ \ \ \ \ \ \ \ \ \ 1st order & $T_c\downarrow$\ \ \ crossover & $T_c\downarrow$\ \ \ crossover \\
\hline
$\Omega$ & $T_c\downarrow$ \ \ \ \ \ \ \ \ \ \ 1st order & $T_c\downarrow$\ \ \ crossover & $/$ \ \ \ \ \ \ \ \ \ \ \ \ \ \ \ \ \ \ \ \ \ \ \ \ \ $/$\\
\hline
\end{tabular}
\caption{A summary of the effects of imaginary ($\Omega_I$) and real ($\Omega$) rotations, where $\uparrow/\downarrow$ stands for increasing/decreasing and the orders of deconfinement transition are also listed.}\label{V12}
\end{table}

%%%%%%%%%%%%%%%%%%%%%%%%%%%%%%%%%%%%%%%%%%%%%%%%%%%%%%%%%%%%%%%%%%%%%%%%%%%%%%%%%%%%%%%%%%%%%%%%%%
\section{The three-flavor Polyakov--Nambu--Jona-Lasinio model}\label{3PNJL}
%%%%%%%%%%%%%%%%%%%%%%%%%%%%%%%%%%%%%%%%%%%%%%%%%%%%%%%%%%%%%%%%%%%%%%%%%%%%%%%%%%%%%%%%%%%%%%%%%%
Now, we extend the study to a full QCD matter with gluon degrees of freedom through the modified Polyakov loop potential and three flavors of quarks through Nambu--Jona-Lasinio (NJL) model~\cite{Klevansky:1992qe,Hatsuda:1994pi}. According to the studies in Sec.\ref{PSGT}, the PL potential $V_2$ are more consistent with the lattice simulations, so we will adopt it for further explorations in this section. On the other hand, the rotation effects on quark dynamics have been explored for many years in NJL model~\cite{Jiang:2016wvv,Chen:2015hfc,Ebihara:2016fwa,Cao:2019ctl,Chen:2019tcp,Zhang:2018ome,Cao:2020pmm,Wang:2018zrn}, so we are now armed with the three-flavor Polyakov--Nambu--Jona-Lasinio (PNJL) model where rotation effects are taken into account in both quark and gluon sectors. Note that unlike the baryon or isospin chemical potential, the rotations can have direct effects on both gluons and quarks though it functions through effective chemical potentials. However, one should note that there is no interplay between the rotation effects of the two sectors except implicitly through the quark--PL coupling. As a consequence, the model might not reproduce the numerical results of LQCD so well but could definitely help us to understand the features qualitatively and physically.

%%%%%%%%%%%%%%%%%%%%%%%%%%%%%%%%%%%%%%%%%%%%%%%%%%%%%%%%%%%%%%%%%%%%%%%%%%%%%%%%%%%%%%%%%%%%%%%%%%
\subsection{The formalism}
%%%%%%%%%%%%%%%%%%%%%%%%%%%%%%%%%%%%%%%%%%%%%%%%%%%%%%%%%%%%%%%%%%%%%%%%%%%%%%%%%%%%%%%%%%%%%%%%%%
By introducing the effect of imaginary rotation, the Lagrangian density of three-flavor PNJL model can be given as~\cite{Fukushima:2017csk,Klevansky:1992qe,Hatsuda:1994pi}:
\begin{eqnarray}
{\cal L}_{\rm PNJL}&=&\bar\psi\left[i\slashed{\partial}-i\gamma^4\left(ig{\cal A}_4+i\,\Omega_{\rm I}(\hat{L}_z+\hat{S}_z)\right)-m_0\right]\psi\nonumber\\
&&+G\sum_{a=0}^8\left[(\bar\psi\lambda^a\psi)^2+(\bar\psi i\gamma_5\lambda^a\psi)^2\right]+{\cal L}_{\rm tH}\nonumber\\
&&-V_2(\phi_1,\phi_2,\Omega_{\rm I}),\label{PNJL}
\end{eqnarray}
where $\psi=(u,d,s)^T$ is the three-flavor quark field with the current mass matrix
 \bea
 m_0&\equiv&{\rm diag}(m_{\rm 0u},m_{\rm 0d},m_{\rm 0s}),
 \eea
 and $g{\cal A}_4$ represents the background $SU(3)$ gauge field. In the imaginary rotation term of quarks, $\hat{L}_z\equiv -i(x_1\partial_2-x_2\partial_1)$ and $\hat{S}_z\equiv{i\over2}\gamma^1\gamma^2$ are the operators of orbital and spin angular momenta, respectively. In the four-quark interaction terms, the vertex $\lambda^0=\sqrt{2/3}~\mathbbm{1}_3$, and $\lambda^i~(i=1,\dots,8)$ are Gell-Mann matrices in flavor space. For later use, the 't Hooft term ${\cal L}_{\rm tH}\equiv-K\sum_{t=\pm}{\rm Det}~\bar\psi\Gamma^t\psi$ can be re-expressed as~\cite{Klevansky:1992qe}
\bea
\!\!\!\!\!\!\!\!{\cal L}_{\rm tH}\!=\!-{K\over2}\sum_{t=\pm}\epsilon_{ijk}\epsilon_{imn}(\bar{\psi}^i\Gamma^t{\psi}^i)(\bar{\psi}^j\Gamma^t{\psi}^m)(\bar{\psi}^k\Gamma^t{\psi}^n)
\eea
with the interaction vertices $\Gamma^\pm=\mathbbm{1}_4\pm\gamma_5$ for right- and left-handed channels, respectively. Here, one should note the Einstein summation convention for the flavor indices $i,j,k,m,n$ and the correspondences between $1,2,3$ and $u,d,s$. 

For this setup, only $g{\cal A}_4$ and the scalar condensates $\sigma_{\rm f}=\langle\bar\psi_{\rm f}\psi_{\rm f}\rangle\ (f=u,d,s)$ are assumed to be nonzero. To facilitate the study, we would like first to reduce ${\cal L}_{\rm tH}$ to an effective form with only four-fermion interactions. By applying the Hartree approximation to contract a pair of quark and antiquark in each term~\cite{Klevansky:1992qe}, we immediately find
\begin{eqnarray}
{\cal L}_{\rm tH}^4
=-{K}\epsilon_{ijk}\epsilon_{imn}\sigma_i\!\left(\bar{\psi}^j{\psi}^m\bar{\psi}^k{\psi}^n
-\bar{\psi}^ji\gamma^5{\psi}^m\bar{\psi}^ki\gamma^5{\psi}^n\right).\nonumber\\\label{LNJL4}
\end{eqnarray}
Now, the Lagrangian of PNJL model Eq.\eqref{PNJL} only effectively involves four-quark interactions after substituting ${\cal L}_{\rm tH}$ by ${\cal L}_{\rm tH}^4$. Again, contracting the quark-antiquark pairs in all the four-quark terms, we find the quark bilinear in the following form
\begin{eqnarray}\label{LNJL2}
\!\!\!\!\!\!\!{\cal L}_{\rm PNJL}^2\!\!=\!\bar\psi\left[i\slashed{\partial}\!-\! i\gamma^4\!\!\left(ig{\cal A}_4+i\,\Omega_{\rm I}(\hat{L}_z+\hat{S}_z)\right)\!-\!m\right]\psi,
\end{eqnarray}
where the dynamical mass matrix is 
\bea
m&=&{\rm diag}(m_{\rm u},m_{\rm d},m_{\rm s}) ,\\
m_{\rm f}&=&m_{\rm 0f}-4G\sigma_{\rm f}+2K\sigma_j\sigma_k\label{mass}
\end{eqnarray}
with $f\neq j\neq k$. The $G$ and $K$ dependent terms in Eq.~\eqref{mass} are from the $U_A(1)$ symmetric and anomalous interactions, respectively. 

Now, the most important mission is to evaluate the bilinear contribution to the thermodynamic potential in the presence of rotation. To do that, we consider a cylindrical system with radius $R$ and simply set the condensates to be homogeneous across the space. Then, the eigenfunction $\psi_{\rm f}\ (f=u,d,s)$ can be presented on the basis of eigenstates of $\hat{L}_z$ and transverse helicity $\hat{h}_\bot\equiv \gamma^5\gamma^3{\bf \hat{k}_{\bot}\cdot \hat{S}}$ as~\cite{Jiang:2016wvv} 
\bea
u_{\rm f,k_\bot,k_z,l,t}=\sqrt{\epsilon_{\rm f}+m_{\rm f}\over4\epsilon_{\rm f}}e^{i\,k_z x_3}\left(\begin{array}{c}
\tilde{J}_l\\
t\,\tilde{J}_{l+1}\\
{k_z-i\,t\,k_\bot\over\epsilon_{\rm f}+m_{\rm f}}\tilde{J}_l\\
{i\,k_\bot-t\,k_z\over\epsilon_{\rm f}+m_{\rm f}}\tilde{J}_{l+1}\\
\end{array}\right)\\
\upsilon_{\rm f,k_\bot,k_z,l,t}=\sqrt{\epsilon_{\rm f}+m_{\rm f}\over4\epsilon_{\rm f}}e^{-i\,k_z x_3}\left(\begin{array}{c}
{k_z-i\,t\,k_\bot\over\epsilon_{\rm f}+m_{\rm f}}\tilde{J}_l\\
{t\,k_z-i\,k_\bot\over\epsilon_{\rm f}+m_{\rm f}}\tilde{J}_{l+1}\\
\tilde{J}_l\\
-t\,\tilde{J}_{l+1}\\
\end{array}\right)
\eea
with $t=\pm$, the energy $\epsilon_{\rm f}(k_\bot,k_3)=\sqrt{k_\bot^2+k_3^2+m_{\rm f}^2}$ and $\tilde{J}_l(k_\bot r,\theta)\equiv e^{i\,l\theta}J_l(k_\bot r)$. So, the quark Feynman propagators are
\bea
G_{\rm f}=\sum_{t=\pm}u_{\rm f,k_\bot,k_z,l,t}\bar{u}_{\rm f,k_\bot,k_z,l,t}
\eea
on the variable space $k_\bot,k_z$ and $l$; and then functional integrations of the quark fields can be carried out to give~\cite{Jiang:2016wvv} 
\begin{widetext}
\bea
\!\!\!\!\Omega_{\rm bl}\!=\!-\!\!\sum_{\rm f=u,d,s}\sum_{l=-\infty}^\infty\int{d k_\bot ^2\over2\pi}\!\!\int\!\!{\di k_3\over2\pi}\int_0^R\!\!{r\di r\over R^2}\left[J_l^2(k_\bot r)\!+\!J_{l+1}^2(k_\bot r)\right]\left\{N_{\rm c}\epsilon_{\rm f}\!+\!{T}\sum_{\rm j=1}^3\sum_{u=\pm}\ln\left[1\!+\!e^{-\tilde{\epsilon}_{\rm f}+u\,i\,\left(q_j+(l+{1\over2})\tilde{\Omega}_{\rm I}\right)}\right]\right\}\label{omgbl}
\eea
with $\tilde{\epsilon}_{\rm f}\equiv\epsilon_{\rm f}/T$. For $\tilde{\Omega}_{\rm I}=0$, recall the normalization property $\sum_{l=-\infty}^\infty J_l^2(k_\bot r)=1$, the thermodynamic potential can be reduced to
\bea
\Omega_{\rm bl}\!=\!-2\!\!\sum_{\rm f=u,d,s}\!\!\int\!\!{dk_\bot^2\di k_3\over8\pi^2}\left\{N_{\rm c}\epsilon_{\rm f}\!+\!{T}\sum_{\rm j=1}^3\sum_{u=\pm}\ln\left[1\!+\!e^{-\tilde{\epsilon}_{\rm f}+u\,i\,q_j}\right]\right\}.\label{omg0}
\eea
This is exactly the same as the one given in Ref.~\cite{Fukushima:2017csk} except that the integration variables are now taken as the transversal momentum $k_\bot$ and longitudinal momentum $k_3$.

For a finite rotation $\Omega$, the radius should satisfy $R\leq \Omega^{-1}$ for the sake of causality. Then, to average the Bessel functions over transversal space, the boundary condition has to be applied at finite $R$. In Sec.\ref{RE}, we have mentioned how boundary condition can be applied to the bosonic fields; here, for a Dirac field, the boundary condition is a little tricky since we cannot require all components of the wave-function to vanish. Instead, no net current across the boundary was adopted as the physical boundary condition~\cite{Ebihara:2016fwa}, and we simply have 
\bea
k_{l,n}R=\left\{\begin{array}{l}
\xi_{l,n},\ \ \ \ \ \ l\geq0\\
\xi_{-l-1,n},\ l<0
\end{array},\right.
\eea
where $\xi_{l,n}$ is the $n$-th zero of $J_l(z)$. So by altering $k_\bot$ to $k_{l,n}$ and the integration to summation correspondingly, the bilinear term Eq.\eqref{omgbl} becomes
\bea
\!\!\!\!\!\!\!\Omega_{\rm bl}\!=\!-\!\!\!\!\sum_{\rm f=u,d,s}\sum_{l=-\infty}^\infty\sum_{\rm n=1}^\infty{2/\pi\over R^2J_{l+1}^2(k_{l,n} R)}\!\!\int\!\!{\di k_3\over2\pi}\int_0^R\!\!{r\di r\over R^2}\left[J_l^2(k_{l,n} r)+J_{l+1}^2(k_{l,n} r)\right]\left\{N_{\rm c}\epsilon_{\rm f}\!+\!{T}\sum_{\rm j=1}^3\sum_{u=\pm}\ln\left[1\!+\!e^{-\tilde{\epsilon}_{\rm f}+u\,i\,q_j}\right]\right\}
\eea
with the weight for the summation over $n$ follows that given in Ref.~\cite{Ebihara:2016fwa}. Recalling the properties $\int_0^R{2r\di r\over R^2}J_l^2(k_{l,n} r)=\int_0^R{2r\di r\over R^2}J_{l+1}^2(k_{l,n} r)=J_{l+1}^2(k_{l,n} R)$, it follows
\bea
\Omega_{\rm bl}\!&=&\!-2\!\!\sum_{\rm f=u,d,s}\sum_{l=-\infty}^\infty{1\over\pi R^2}\sum_{\rm n=1}^\infty\int\!\!{\di k_3\over2\pi}\left\{N_{\rm c}\epsilon_{\rm f}\!+\!{T}\sum_{\rm j=1}^3\sum_{u=\pm}\ln\left[1\!+\!e^{-\tilde{\epsilon}_{\rm f}+u\,i\,\left(q_j+(l+{1\over2})\tilde{\Omega}_{\rm I}\right)}\right]\right\}\nonumber\\
\!&=&\!-2\!\!\sum_{\rm f=u,d,s}\sum_{l=0}^\infty{1\over\pi R^2}\sum_{\rm n=1}^\infty\int\!\!{\di k_3\over2\pi}\left\{2N_{\rm c}\epsilon_{\rm f}\!+\!{T}\sum_{\rm j=1}^3\sum_{u,t=\pm}\ln\left[1\!+\!e^{-\tilde{\epsilon}_{\rm f}+u\,i\,\left(q_j+t(l+{1\over2})\tilde{\Omega}_{\rm I}\right)}\right]\right\}.\label{omgR}
\eea
Here, it is interesting to notice that ${1\over\pi R^2}$ plays a role of degeneracy factor for the summation over $n$ and the corresponding eigenenergies satisfy $k_{l,n}^2\propto R^{-2}$. This is quite similar to the case with a finite magnetic field~\cite{Cao:2021rwx}: the degeneracy factor is ${|qB|\over2\pi}$ for the higher Landau levels $n\geq1$, and the corresponding Landau energies are $2n|qB|$. However, the effects of finite size and magnetic field are completely different due to the Lowest Landau level, with a vanishing energy, in a magnetic field~\cite{Cao:2021rwx}.

In the limit $\Omega_{\rm I}\rightarrow 0$, we find that the bilinear potential $\Omega_{\rm bl}$ in Eq.\eqref{omgR} is still $R$-dependent, inconsistent with Eq.\eqref{omg0}. The reason is that the boundary condition has not been self-consistently taken into account in Eq.\eqref{omg0} and the form only corresponds the thermodynamic limit $R\rightarrow\infty$ where the boundary effect is negligible. In this sense, Eq.\eqref{omgR} must be consistent with Eq.\eqref{omg0} in the limit $R\rightarrow\infty$ as the system considered is exactly the same. For $\Omega_{\rm I}=0, T=0.15\,{\rm GeV}$ and $R=800\,{\rm GeV}^{-1}$, we have checked numerically that the convergent thermal parts are consistent with each other within an error $0.1\%$. On the other hand, the vacuum part of the thermodynamic potential is divergent and can be regularized as
\bea
\Omega_{\rm bl}&=&-2\!\!\sum_{\rm f=u,d,s}\sum_{l=0}^\infty{1\over\pi R^2}\sum_{\rm n=1}^\infty\int\!\!{\di k_3\over2\pi}\left\{2N_{\rm c}\epsilon_{\rm f}^\Lambda+{T}\sum_{\rm j=1}^3\sum_{u,t=\pm}\ln\left[1+e^{-\tilde{\epsilon}_{\rm f}+u\,i\,\left(q_j+t(l+{1\over2})\tilde{\Omega}_{\rm I}\right)}\right]\right\}
\eea
by adopting the Pauli-Villars scheme, where the regularized vacuum energy is~\cite{Klevansky:1992qe}
\bea
\epsilon_{\rm f}^\Lambda=\sum_{j=0}^3(-1)^jC_3^{j}\sqrt{\epsilon_{\rm f}^2+j\,\Lambda^2}.
\eea
To facilitate numerical calculations, the integration over $k_3$ can be carried out for the vacuum part and the thermal part can be rewritten by utilizing the condition $q_1+q_2+q_3=0$, we have
\bea
\Omega_{\rm bl}&=&-\!\!\sum_{\rm f=u,d,s}\sum_{l=0}^\infty{1\over\pi^2R^2}\sum_{\rm n=1}^\infty\!\!\Bigg\{{N_{\rm c}}\sum_{j=0}^3(-1)^{j-1}C_3^{j}(\epsilon_{\rm f0}^2+j\Lambda^2)\ln(\epsilon_{\rm f0}^2+j\Lambda^2)\nonumber\\
&&+2{T}\sum_{t=\pm}\int_0^\infty{\di k_3}\left[\ln\left(1+3Le^{-\tilde{\epsilon}_{\rm f}+i\,t(l+{1\over2})\tilde{\Omega}_{\rm I}}+3L^*e^{-2\tilde{\epsilon}_{\rm f}+2i\,t(l+{1\over2})\tilde{\Omega}_{\rm I}}+e^{-3\tilde{\epsilon}_{\rm f}+3i\,t(l+{1\over2})\tilde{\Omega}_{\rm I}}\right)+c.c.\right]\Bigg\}\label{omgL}
\eea
with $\epsilon_{\rm f0}\equiv\epsilon_{\rm f}(k_{l,n},0)$ and $C_2^{-1}=0$.

Eventually, the coupled gap equations of chiral condensates follow directly from the definitions
$\sigma_{\rm f}\equiv\langle\bar{q}_{\rm i}{q}_{\rm i}\rangle={\partial\Omega_{\rm bl}\over\partial m_{\rm f}}$ as~\cite{Klevansky:1992qe}
\bea
\sigma_{\rm f}
&=&-{2\over\pi^2 R^2}\sum_{l=0}^\infty\sum_{\rm n=1}^\infty \Bigg\{{N_{\rm c}}m_{\rm f}\sum_{j=0}^3(-1)^{j-1}C_3^{j}\ln(\epsilon_{\rm f0}^2+j\Lambda^2)\nonumber\\
&&-3\sum_{t=\pm}\int_0^\infty{\di k_3}{m_{\rm f}\over\epsilon_{\rm f}}\left[{Le^{-\tilde{\epsilon}_{\rm f}+i\,t(l+{1\over2})\tilde{\Omega}_{\rm I}}+2L^*e^{-2\tilde{\epsilon}_{\rm f}+2i\,t(l+{1\over2})\tilde{\Omega}_{\rm I}}+e^{-3\tilde{\epsilon}_{\rm f}+3i\,t(l+{1\over2})\tilde{\Omega}_{\rm I}}\over1+3Le^{-\tilde{\epsilon}_{\rm f}+i\,t(l+{1\over2})\tilde{\Omega}_{\rm I}}+3L^*e^{-2\tilde{\epsilon}_{\rm f}+2i\,t(l+{1\over2})\tilde{\Omega}_{\rm I}}+e^{-3\tilde{\epsilon}_{\rm f}+3i\,t(l+{1\over2})\tilde{\Omega}_{\rm I}}}+c.c.\right]\Bigg\}.\label{gapf}
\eea
And the total self-consistent thermodynamic potential of the PNJL model can be found to be~\cite{Cao:2021gfk}
\bea
\Omega_{\rm PNJL}&=&2G\sum_{\rm f=u,d,s}\sigma_{\rm f}^2-4K\sigma_{\rm u}\sigma_{\rm d}\sigma_{\rm s}+\Omega_{\rm bl}+V_2(\phi_1,\phi_2,\Omega_{\rm I})\label{OPNJL}
\eea
by utilizing the definitions of chiral condensates and their relations to dynamical masses Eq.\eqref{mass}. Since $\Omega_{\rm bl}$ also depends on the background gauge fields $q_{\rm i}\ (i=1,2,3)$, the gap equations of $\phi_1$ and $\phi_2$ follow $\partial\Omega_{\rm PNJL}/\partial{\phi_i}=0\ (i=1,2)$ as
\bea
0&=&-6T\sum_{\rm f=u,d,s}{1\over\pi R^2}\sum_{l=0}^\infty\sum_{\rm n=1}^\infty\int{\di k_3\over2\pi}\sum_{t=\pm}\left[{{\partial L\over\partial{\phi_i}}e^{-\tilde{\epsilon}_{\rm f}+i\,t(l+{1\over2})\tilde{\Omega}_{\rm I}}+{\partial L^*\over\partial{\phi_i}}e^{-2\tilde{\epsilon}_{\rm f}+2i\,t(l+{1\over2})\tilde{\Omega}_{\rm I}}\over1+3Le^{-\tilde{\epsilon}_{\rm f}+i\,t(l+{1\over2})\tilde{\Omega}_{\rm I}}+3L^*e^{-2\tilde{\epsilon}_{\rm f}+2i\,t(l+{1\over2})\tilde{\Omega}_{\rm I}}+e^{-3\tilde{\epsilon}_{\rm f}+3i\,t(l+{1\over2})\tilde{\Omega}_{\rm I}}}+c.c.\right]\nonumber\\
&&+{\partial V_2(\phi_1,\phi_2,\Omega_{\rm I})\over\partial{\phi_i}}\label{gapi}
\eea
with the last term given by the right-hand sides of Eqs.\eqref{dV11} and \eqref{dV12}. And according to Eqs.\eqref{PL} and \eqref{qi}, the derivatives of $L$ can be evaluated as
\bea
{\partial L\over\partial{\phi_1}}={i\over6}\left(-e^{i\,q_1}+e^{i\,q_3}\right),\ {\partial L\over\partial{\phi_2}}={i\over6\sqrt{3}}\left(-e^{i\,q_1}+2e^{i\,q_2}-e^{i\,q_3}\right).
\eea 

For the case with real rotation, the thermodynamic potential and gap equations can be directly obtained from Eqs.(\ref{omgL}-\ref{gapi}) by taking the analytic continuation: $\Omega_{\rm I}\rightarrow-i\,\Omega$. At first glance, it seems that the thermal part in Eq.\eqref{omgL} would diverge with increasing $l$. In fact, it would not if one notices that the eigenenergy $k_{l,n}$ is also $l$ dependent: Since $\xi_{l+1,n}>\xi_{l,n}+1$ and $\xi_{0,1}\approx2.4$, we find $k_{l,n}>(l+1/2)R^{-1}>(l+1/2){\Omega}$ after applying the causality condition ${\Omega}R\leq 1$ and the divergence disaster is well avoided. We want to emphasize that by presenting the OAM $l$ in the form of a natural number, the imaginary parts cancel out in the thermodynamic potential of quarks even for a real rotation, which means that there is no sign problem at all for the LQCD simulations~\cite{Fukushima:2017csk}. To achieve that, it is important that only homogeneous phases are considered and the boundary condition is self-consistently taken into account. 
\end{widetext}

%%%%%%%%%%%%%%%%%%%%%%%%%%%%%%%%%%%%%%%%%%%%%%%%%%%%%%%%%%%%%%%%%%%%%%%%%%%%%%%%%%%%%%%%%%%%%%%%%%
\subsection{Numerical results}
%%%%%%%%%%%%%%%%%%%%%%%%%%%%%%%%%%%%%%%%%%%%%%%%%%%%%%%%%%%%%%%%%%%%%%%%%%%%%%%%%%%%%%%%%%%%%%%%%%
Now, we are going to solve the five coupled gap equations for given $T$ and $\Omega_{\rm I}$, that is, three in Eq.\eqref{gapf} and two in Eq.\eqref{gapi}. The model parameters can be fixed as
\bea
&&m_{0u}=m_{0d}=5.28~{\rm MeV}, m_{0s}=0.124~{\rm GeV}, \nonumber\\
&&\Lambda\!=\!1.13~{\rm GeV}, G\!=\!3.51~{\rm GeV}^{-2}, K\!=\! -6.50~{\rm GeV}^{-5}\label{parameters}
\eea
by best fitting to the chiral condensates, pion decay constant, and masses of pion, kaon and $\eta$ mesons from the Particle Data Group, that is,
\bea
&&\!\!\!\!\!\langle\bar{u}u\rangle=\langle\bar{d}d\rangle=(-0.25~{\rm GeV})^3,\ f_\pi = 93~{\rm MeV},\nonumber\\
&&\!\!\!\!\!m_\pi = 138~{\rm MeV},\ m_{K}= 496~{\rm MeV}, m_{\eta}=548~{\rm MeV}.
\eea
Note that we are not able to reproduce the kaon and $\eta$ masses simultaneously with the Pauli-Villars regularization. So the model parameters $m_{0s}$ and $K$ are fixed by giving the smallest deviations from the experimental masses, that is, $(\bar{m}_{K}-m_{K})^2+(\bar{m}_{\eta}-m_{\eta})^2$. For the parameters in Eq.\eqref{parameters}, it turns out that the best fitting masses are $\bar{m}_{K}= 0.409~{\rm GeV}$ and $\bar{m}_{\eta}=0.600~{\rm GeV}$, respectively. Moreover, the dynamical masses of quarks are $m_{\rm u}^v=m_{\rm d}^v=0.220~{\rm GeV}$ and $m_{\rm s}^v=0.454~{\rm GeV}$ in vacuum, which are all smaller than the ones with hard-cutoff regularization~\cite{Cao:2021gfk}. 

\begin{figure}[!htb]
	\begin{center}
	\includegraphics[width=8cm]{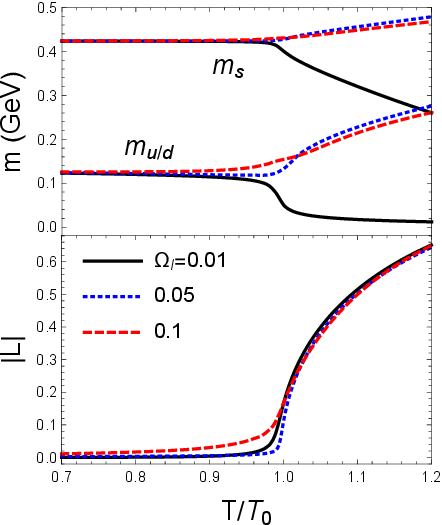}
		\caption{The masses $m_{\rm f}~(f=u,d,s)$ (upper panel) and Polyakov loop $|L|$ (lower panel) as functions of temperature $T$ for imaginary rotations $\Omega_{\rm I}=0.01~{\rm GeV}$ (black solid lines), $0.05~{\rm GeV}$ (blue dotted lines) and $0.1~{\rm GeV}$ (red dashed lines).}\label{mLI}
	\end{center}
\end{figure}
\begin{figure}[!htb]
	\begin{center}
	\includegraphics[width=8cm]{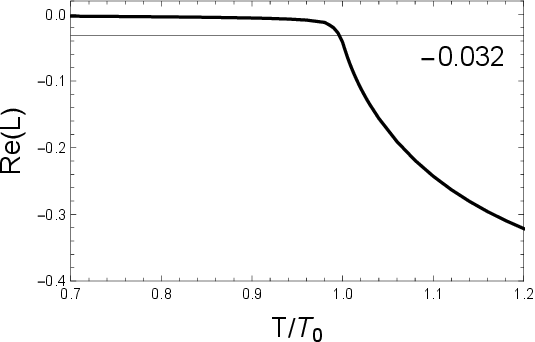}
		\caption{The real part of the Polyakov loop $Re(L)$ as a function of temperature with the baseline corresponding to $Re(L)=-0.032$. }\label{ReL}
	\end{center}
\end{figure}

In the following, we take a small system with radius $R=10~{\rm GeV}^{-1}$ for example, so that the causality condition allows a rotation velocity as large as $0.1~{\rm GeV}$. For imaginary rotation, the numerical results are illustrated in Fig.\ref{mLI}, which are found to be qualitatively consistent with the LQCD simulations~\cite{Braguta:2021jgn,Yang:2023vsw}. Though the absolute value of the Polyakov loop, $|L|$, increases with temperature regardless of the value of $\Omega_{\rm I}$, the features of the dynamical masses or chiral condensates are quite nontrivial: For small $\Omega_{\rm I}$, $m_{\rm f}~(f=u,d,s)$ decrease with $T$; but for large $\Omega_{\rm I}$, $m_{\rm f}~(f=u,d,s)$ slightly decrease and then increase with $T$. The latter feature is unexpected as temperature usually tends to restore chiral symmetry thus reduce dynamical masses~\cite{Klevansky:1992qe,Hatsuda:1994pi}, but is consistent with the LQCD results~\cite{Yang:2023vsw}. The reason of the anomalous behavior is that the real part of the PL, $Re(L)$, is negative for larger $\Omega_{\rm I}$ and $T$, contrary to the positive value in general case. As a consequence, the temperature effect can be reversed by the couplings between $L/L^*$ and one/two quark terms in Eq.\eqref{omgL} when 
\bea
Re\left[3Le^{-\tilde{\epsilon}_{\rm f}+i\,{t\over2}\tilde{\Omega}_{\rm I}}+3L^*e^{-2\tilde{\epsilon}_{\rm f}+2i\,{t\over2}\tilde{\Omega}_{\rm I}}+e^{-3\tilde{\epsilon}_{\rm f}+3i\,{t\over2}\tilde{\Omega}_{\rm I}}\right]<0.\nonumber\\
\eea
For $\Omega_{\rm I}=0.05~{\rm GeV}$, we can take $\tilde{\Omega}_{\rm I}=0$ around $T_0$, then by adopting the lowest transversal energy $k_{0,1}$ in $\tilde{\epsilon}_{\rm f}$, the inequation implies that $Re(L)<-0.032$. This is roughly consistent with the threshold of increasing masses in the upper panel of Fig.\ref{mLI}, see Fig.\ref{ReL}.
Furthermore, according to the results in Fig.\ref{mLI}, the transition becomes crossover when quark degrees of freedom are taken into account; but contrary to the LQCD simulations~\cite{Yang:2023vsw}, the pseudocritical temperature increases with $\Omega_{\rm I}$ (though very little). The latter observation indicates that the chiral symmetry breaking effect of $\Omega_{\rm I}$ is stronger than its deconfinement effect in the modified PNJL model. To reproduce the LQCD results better, more realistic PL potential and quark-gluon interplay are needed in the PNJL model. 

\begin{figure}[!htb]
	\begin{center}
	\includegraphics[width=8cm]{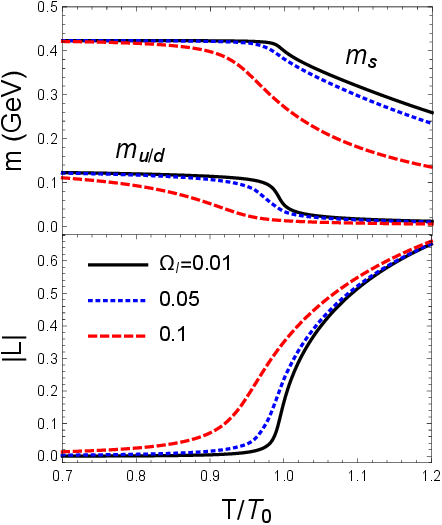}
		\caption{The masses $m_{\rm f}~(f=u,d,s)$ (upper panel) and Polyakov loop $|L|$ (lower panel) as functions of temperature $T$ for real rotations $\Omega=0$ (black solid lines), $0.05~{\rm GeV}$ (blue dotted lines) and $0.1~{\rm GeV}$ (red dashed lines).}\label{mLR}
	\end{center}
\end{figure}
\begin{figure}[!htb]
	\begin{center}
	\includegraphics[width=8cm]{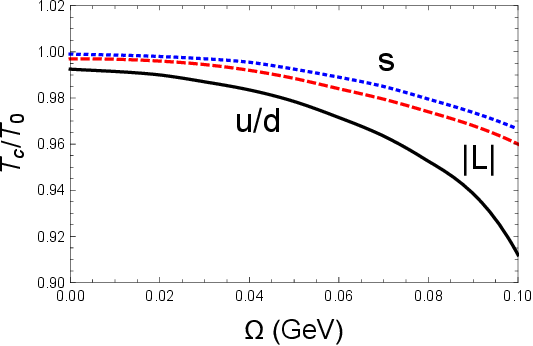}
		\caption{The pseudocritical temperatures $T_c$ as functions of real rotation $\Omega$ for $m_{\rm u/d},\ m_{\rm s}$ and $|L|$ in the PNJL model.}\label{Tomg}
	\end{center}
\end{figure}

For the real rotation, the results are illustrated in Fig.\ref{mLR}. The features of $m_{\rm f}~(f=u,d,s)$ and $|L|$ are all consistent with the initial expectations, that is, $|L|$ increases with both $T$ and $\Omega$ due to deconfinement while $m_{\rm f}~(f=u,d,s)$ decrease with both $T$ and $\Omega$ due to chiral symmetry restoration. The corresponding phase diagram is presented in Fig.\ref{Tomg}, where the pseudocritical temperatures $T_c$, defined by the peak of the absolute values of the susceptibilities, decrease with $\Omega$ as should be~\cite{Jiang:2016wvv}. Note that the $T_c$ of the PL $|L|$ is closer to that of $s$ quark dynamical mass $m_{\rm s}$. Again, similar to the pure gauge theory, such effects of real rotation are opposite to those analytically continued from the results of the imaginary rotation~\cite{Braguta:2021jgn,Yang:2023vsw}, so the model explorations in Refs.~\cite{Chen:2022mhf,Chen:2023cjt} are not necessarily "unsuccessful". We hope that the LQCD simulations with real rotation can help to settle the issues in the future as we have shown that there is no sign problem for a special consideration.

%%%%%%%%%%%%%%%%%%%%%%%%%%%%%%%%%%%%%%%%%%%%%%%%%%%%%%%%%%%%%%%%%%%%%%%%%%%%%%%%%%%%%%%%%%%%%%%%%%
\section{summary}\label{sum}
%%%%%%%%%%%%%%%%%%%%%%%%%%%%%%%%%%%%%%%%%%%%%%%%%%%%%%%%%%%%%%%%%%%%%%%%%%%%%%%%%%%%%%%%%%%%%%%%%%
The work is composed of two parts: the pure $SU(3)$ gauge theory and the Polyakov--Nambu-Jona-Lasinio model. Firstly, we try to introduce the rotational effect to the pure $SU(3)$ gauge theory inspired from the perturbative study~\cite{Chen:2022smf} but without introducing any extra free parameters. Then, the rotational effects on confinement are compared for the two most popular empirical Polyakov loop potentials given by K. Fukushima and Munich's group, respectively. Secondly, the more successful Munich's potential is applied to the chiral effective PNJL model in order to explore the features of chiral symmetry and confinement simultaneously for a QCD matter.

For the first part, the main findings can be summarized as the following: For the PL potential of Fukushima, a smaller imaginary rotation $\Omega_{\rm I}$ tends to suppress PL at all temperature and the deconfinement transition keeps of first order. At larger $\Omega_{\rm I}$, two more branches of transitions can be identified: $|L|\approx0.8\rightarrow |L|\lesssim 1$ of first order and $|L|\lesssim 1\rightarrow |L|=1$ of second order, both the critical temperatures of which decrease with $\Omega_{\rm I}$. However, for the PL potential of Munich's group, $\Omega_{\rm I}$ tends to enhance PL at low temperature $T$, consistent with lattice simulations~\cite{Braguta:2021jgn,Yang:2023vsw}; but suppress PL at high $T$, consistent with perturbative calculations~\cite{Chen:2022smf}. Moreover, we only find one branch of deconfinement transition which alters from first order to crossover with increasing $\Omega_{\rm I}$, expected from lattice simulations~\cite{Braguta:2021jgn}. On the other hand, the PL is consistently enhanced by the real rotation $\Omega$ at relatively low $T$ in both potentials, and the (pseudo-)critical temperature $T_{\rm c}$ decreases with $\Omega$. Therefore, we at least find an example that the phase diagrams $T-\Omega_{\rm I}$ and $T-\Omega$ cannot be analytically continued to each other, though $\Omega_{\rm I}$ is introduced through analytic continuation of $\Omega$, that is, $\Omega\rightarrow i\,\Omega_{\rm I}$~\cite{Braguta:2021jgn,Yang:2023vsw}. The reason is that the PL potential is not a trivial function of $L$ and $L^*$ anymore in the presence of rotations, and any deviation of $L/L^*$ from $0$ would enhance $|L|$.

For the second part, we find that the interplays between quarks and gluons render the pseudocritical temperature $T_{\rm c}$ insensitive to $\Omega_{\rm I}$ as its effects are opposite from the two sectors while $T_{\rm c}$ decreases with $\Omega$ as the effects are the same. For a larger $\Omega_{\rm I}$, we surprisingly find that temperature would further break chiral symmetry as was also presented in the LQCD results~\cite{Yang:2023vsw}. The reason is that $Re(L)$ becomes more negative with increasing $\Omega_{\rm I}$ and the effect of temperature is reversed due to the linear dependences on $L$ and $L^*$ in the logarithmic terms of $\Omega_{\rm bl}$. In the future, we hope the LQCD groups could directly study the real rotation effect by adopting Taylor expansions -- that would better help us to understand what is the real situation with rotation.

\section*{Acknowledgement}
G. Cao thanks Xu-guang Huang and Hao-lei Chen for useful discussions.

\end{document}